# A Method for Greatly Reduced Edge Effects and Crosstalk in CCT Magnets

M. Koratzinos, ETH Zurich, G. Kirby, J. Van Nugteren, CERN, E. R. Bielert, Univ. Illinois at Urbana

*Abstract*— **Iron-free CCT magnet design offers many advantages, one being the excellent field quality and the absence of multipole components. However, edge effects are present, although they tend to integrate out over the length of the magnet. Many modern accelerator applications, however, require that these magnets are placed in an area of rapidly varying optics parameters, so magnets with greatly reduced edge effects have an advantage. We have designed such a magnet (a quadrupole) by adding multipole components of the opposite sign to the edge distortions of the magnet. A possible application could be the final focus magnets of the FCC-ee, where beam sizes at the entry and exit point of the magnets vary by large factors. We have then used this technique to effectively eliminate cross talk between adjacent final focus quadrupoles for the incoming and outgoing beams.**

## I. Introduction

CANTED- cosine-theta (CCT) magnets have been around since the seventies [1], however only recently have they become popular with magnet designers [2] [3], due to the advent of modern manufacturing techniques (CNC machines and 3D printing). The CCT design concept is based around a pair of conductors wound and powered such that their transverse field components sum up and their axial (solenoidal) fields cancel. In practice the conductor is wound on a pre-cut groove in a supporting hollow cylinder or *mandrel*. The area between grooves is referred to as the *rib* and the supporting solid substrate the *spar*. The difference with a conventional design is that stresses cannot accumulate between conductors but instead forces are intercepted by ribs that transfer the stress to the spar.

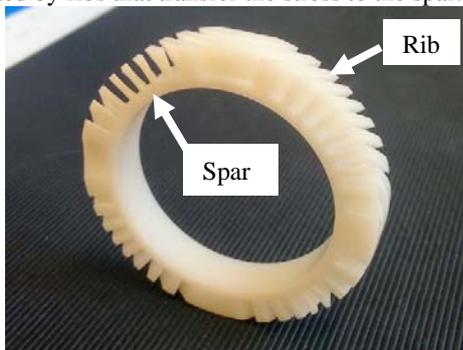

Fig. 1: A section of one of the two layers of a CCT magnet (a quadrupole). The spar is 2 mm wide, followed by a 4 mm rib where the grooves for the cable are located.

In the general case of a coil that produces an arbitrary selection of multipole fields, the centre-line defining the shape of the groove (and the position of the centre of the powered cable) for one of the two coils of the CCT is described by the equation

$$\begin{aligned}x &= R\cos\theta\,;\\ y &= R\sin\theta\,;\\ z &= \sum_{n_B}\left[\frac{R\sin(n_B\theta)}{n_B\tan\alpha_{n_B}}+\frac{\omega\theta}{2\pi}\right]\\ &\quad + \sum_{n_A}\left[\frac{R\cos(n_A\theta)}{n_A\tan\alpha_{n_A}}+\frac{\omega\theta}{2\pi}\right]\end{aligned} \quad (1)$$

Where $R$ is the radius of the coil, A and B are the skew and normal components of the field, $n_A$ and $n_B$ are the skew and normal multipoles ($n_B = 1$ is the dipole component, $n_B = 2$ the quadrupole component, etc., same with $n_A = 1$: skew dipole component, etc.). The angles $\alpha_{n_A}$, which could be a function of $z$, are the angles of the groove (or wire) with respect to the horizontal on the mid plane per desired multipole (called the skew angles). An angle of zero would ensure no relevant multipole component. $\theta$ runs from 0 to $2\pi n_t$ where $n_t$ is the number of turns. For the second layer, $R$ is slightly increased (depending on the thickness of the spar and the cable) and the skew angle has the opposite sign. The start and end of both layers are located on top of one another. We can see from equation (1) that the groove and cable describe a circle in the x-y plane whereas in the longitudinal (z) direction there is a longitudinal shift parameter $\omega$ per revolution, plus the multipole component.

The CCT design offers significant advantages over traditional magnet design for certain applications. Their field quality is excellent due to the purity of the design and due to the fact that the cable grooves can be very precisely machined; they are easy to manufacture using CNC machines or even 3D printing techniques, leading to very fast prototyping; there is no need for coil pre-stress during assembly, leading to simple and fast winding; reduced coil stresses improve magnet training; total freedom to design any multipole arrangement, therefore capable of producing compact double aperture magnets with the required field quality, as demonstrated in this paper; and finally this concept uses fewer components and is considerably lighter than traditional designs, leading to reduced overall costs.

The disadvantage of the CCT design is that the two magnet coils work against each other to cancel the longitudinal field,

leading to more conductor material per Tesla produced. Since our application is not a high field application, we are not affected by this potential limitation in this design.

## II. EDGES CORRECTION

### A. The coil

The coil we are going to use to demonstrate the edges correction is one of the two final focus quadrupoles of FCC-ee [4] [5]. It is a quadrupole of magnetic length of 1200 mm and with an inner bore of diameter 40 mm. The cable of the inner coil has an inner and outer radius of 22 and 26 mm, and the outer coil 28 and 32 mm. the inner spar occupies the area of radius 20 to 22 mm and the middle spar a radius of 26 to 28 mm. The grooves are 2 mm wide and 4 mm high, leaving a possible cross section for the cable of 8 mm$^2$. The pitch between grooves is 5 mm, leaving a minimum rib width of 1mm. The beam pipe is expected to have a diameter of 30 mm, so all multipoles are calculated at a radius of 10 mm, at an aperture of 2/3. This quadrupole produces a gradient of 100 T/m for a total current of around 5800 A. The transverse components of the magnet can be seen in Fig. 2.

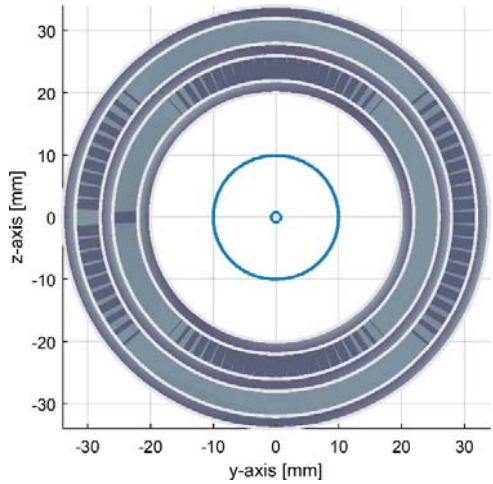

Fig. 2    A view in the transverse plane of the CCT magnet (a quadrupole) used here to demonstrate the edge effect correction. The blue circle (radius 10mm) is where multipoles are calculated.

The software used throughout this analysis is the Field 2017 suite of programs [6].

The multipole components around both edges of such a magnet can be seen in Fig. 3. Only one magnet edge is shown (the one at negative z). The other edge has components which are antisymmetric. All A and B components integrate to zero when integrating over the length of the magnet. However, only the B components integrate to zero locally (per edge). As this magnet will be placed in an area of rapidly varying optics functions, it is beneficial if an edge correction could be applied so that the multipoles would integrate to zero locally.

The integral of the multipole components (normalized to the B2 field, in units of $10^{-4}$) can be seen in Fig. 4.

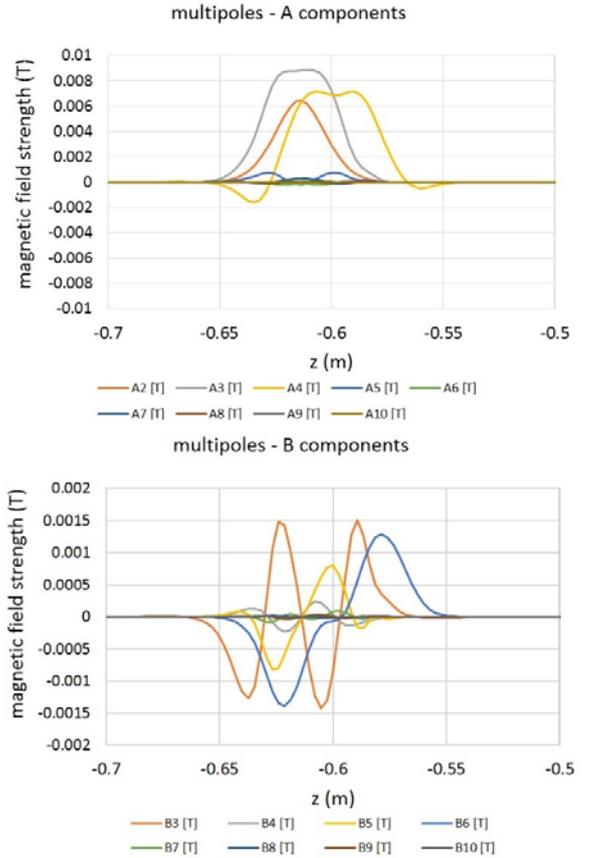

Fig. 3:    A and B multipole components up to order 10 on the left edge of the coil. The A, B1 and B2 components have been omitted. The right edge has components with a flipped sign.

### B. The correction

The correction needs to be applied locally to the A components. This is done by applying non-zero multipole components for the first two turns of the coil. To make sure that the cable does not turn back on itself (i.e. that the gap between the adjacent windings of the cable is always larger than zero) the pitch for these first two windings has been increased to 15mm from 5 mm for the rest of the coil. Corrections up to order 6 are performed (for higher orders the residual effect is too small). Following the A corrections, some B component corrections need to be also applied, again for the first two turns of the coil. This analysis is performed in the absence of any alignment or positioning errors. The integrated multipole plot after the correction can be seen in Fig. 6. This demonstrates that a correction to an arbitrary degree of accuracy can be achieved (here we have stopped the process when an accuracy of 0.05 units or better had been achieved). The magnitude of the edge corrections can be seen in Table 1.

TABLE 1
SIZE OF EDGE CORRECTION (IN DEGREES) FOR THE FIRST AND LAST WINDINGS OF THE MAGNET FOR ALL CORRECTED MULTIPOLES. B2, THE MAIN COMPONENT, IS ALSO GIVEN FOR REFERENCE

|  | A2 | A3 | A4 | A5 | A6 | B2 | B3 | B4 | B5 | B6 |
|---|---|---|---|---|---|---|---|---|---|---|
| $\alpha$ left | -3.1 | 19 | -38 | 6 | 6 | 60 | -5 | -3.5 | 6.5 | 1.5 |
| $\alpha$ right | 3.1 | -19 | 38 | -6 | -6 | 60 | -5 | -3.5 | 6.5 | 1.5 |

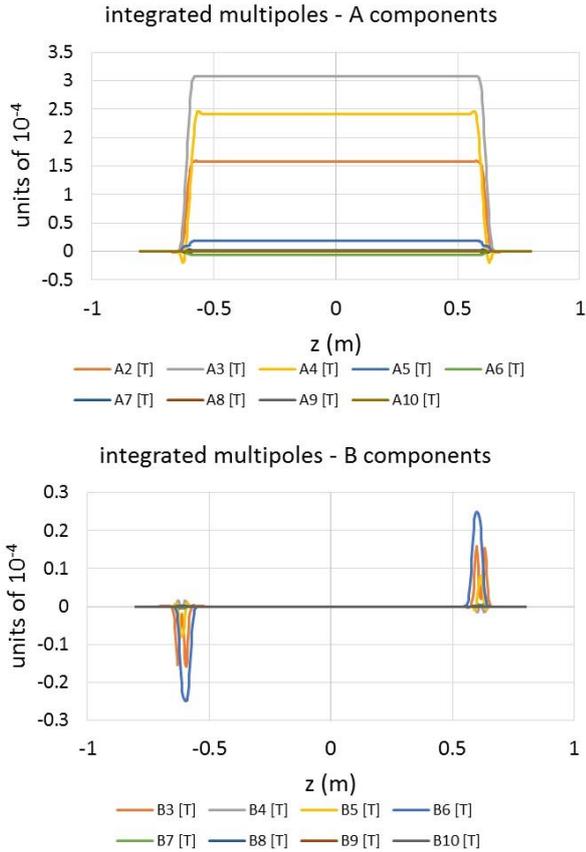

Fig. 4  Integrated multipoles in units of $10^{-4}$. The A1, B1 and B2 components have been left out for clarity. The A1 and B1 components do not need to be corrected, whereas the B2 component has a final integrated value (by definition) of 10,000.

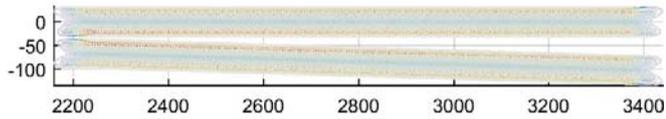

Fig. 5:  The position of the two QC1L1 magnets, at an angle of 30 mrad and at a distance of 2.2 m from the IP. The axes go through the positron beamline.

## III. Crosstalk compensation

### A. The coils

Up to now we have been working with a single coil in standalone mode. However, in the case of the FCC-ee final focus magnets the beam pipes for the electrons and positrons (traveling in opposite directions) intercept at an angle of 30 mrad at the interaction point (IP), in the horizontal plane. The first final focus quadrupole, called QC1L1, starts at a distance of 2.2 m from the IP. There are two quadrupoles, one for the electron and one for the positron beam. Their distance from their magnetic centers is 66 mm at the tip and 102 mm at the end away from the IP (the magnets are 1.2 m long), as can be seen in Fig. 5). The FCC-ee final focus system has many more magnetic elements, but we will concentrate on the crosstalk compensation of the two QC1L1 magnets, which is the most challenging problem. No iron is present in the vicinity of the magnets. The two QC1L1 magnets are a mirror image of each other (the hypothetical mirror standing vertically between the two beamlines).

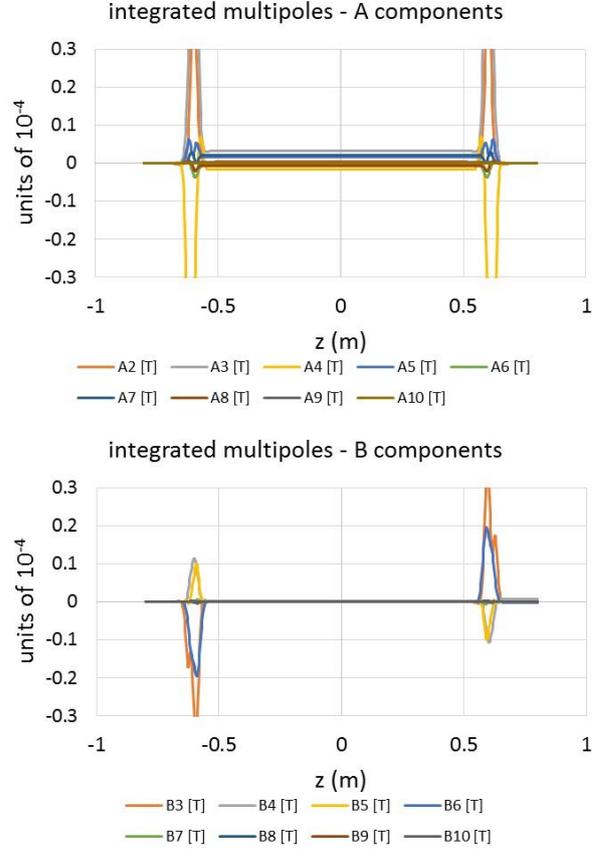

Fig. 6  Integrated multipoles in units of $10^{-4}$ after correction. Note the different scale compared to Fig. 4.

The uncorrected multipoles from this arrangement can be seen in Fig. 7. There are significant components due to the close proximity of the other coil.

### B. The method

Every effort is made to perform any needed correction locally. Currently the correction is performed empirically, with plans to develop an automated minimization procedure in the near future. Multipole corrections are nearly orthogonal to each other, so the minimization process converges rapidly. Only exception is the edges A2 correction which is affected by other multipoles, therefore the correction for A2 should be performed last.

It is not clear where the limits of the method are with respect to the level of compensation possible. We simply stopped at a level (around 0.05 units) where we felt that other distortions (for instance, due to misalignment or winding errors) would be more important.

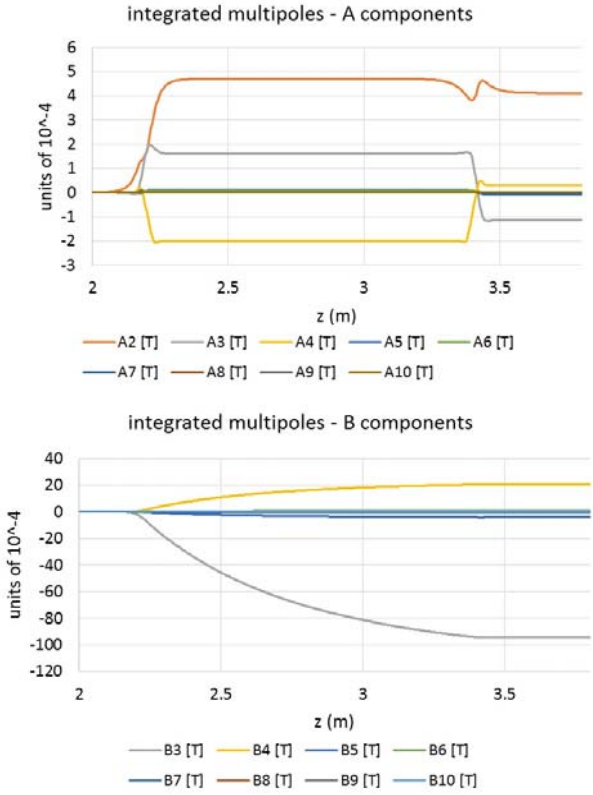

Fig. 7: Integrated multipoles in units of $10^{-4}$ before correction in the case of two side-by-side QC1L1 magnets. As expected from the proximity of the two quadrupoles, the effect of cross talk is large.

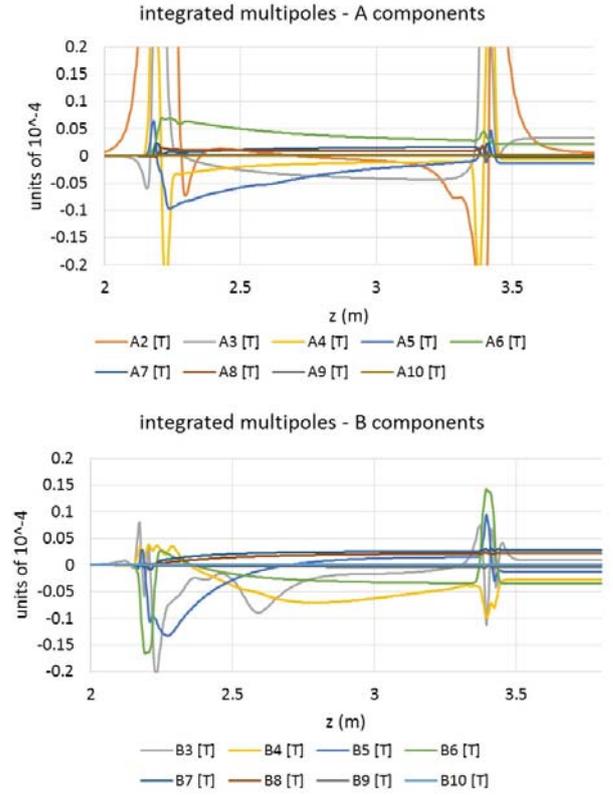

Fig. 8: Integrated multipoles in units of $10^{-4}$ after correction for the effect of crosstalk from the adjacent quadrupole. All multipoles can be corrected to better than 0.05 units

## C. The correction

In contrast to the earlier case, here we need to introduce multipole components along the whole length of the magnet. The results are very encouraging and can be seen in Fig. 8. All multipoles are corrected to within 0.05 units. The maximum and minimum correction along the length of the magnet (excluding the edges where a special correction is performed) can be seen in Table 2.

TABLE 2
SIZE OF CROSSTALK CORRECTION (IN DEGREES) ALONG THE LENGTH OF THE QUADRUPOLE. THE EDGES HAVE BEEN EXCLUDED FROM THIS TABLE. B2, THE MAIN COMPONENT, IS ALSO GIVEN FOR REFERENCE

|   | A2 | A3 | A4 | A5 | A6 | B2 | B3 | B4 | B5 | B6 |
|---|---|---|---|---|---|---|---|---|---|---|
| $\alpha$ max | 0 | 0 | 0 | 0 | 0 | 60 | 5.1 | -4.0 | 2.0 | -1.4 |
| $\alpha$ min | 0 | 0 | 0 | 0 | 0 | 60 | 0.8 | -0.3 | 0.1 | -0.0 |

## IV. CONCLUSION

CCT magnets offer a versatility seldom associated with magnet design. Any multipole arrangement can be designed and implemented. We have first demonstrated that the inevitable edge effects of our test CCT quadrupole magnet (and therefore any CCT magnet) can be eliminated to below 0.05 units. We have further demonstrated that in an iron-free environment we can create two nearly-perfect parallel powered quadrupoles that have a gap of only 2 mm at one tip and 4 cm at the other. Again, the correction is such that residual multipole components can be kept well below 0.05 units. This design eliminates the need for a large number of corrector magnets and might be important in an application where space is very limited and optics performance very important, like the interaction region of the FCC-ee study.

## REFERENCES

[1] D. I. Meyer and R. Flasck, "A new configuration for a dipole magnet for use in high energy physics applications," *NIM A 80.2 (1970), pp 339-341*.

[2] S. Caspi, "Design, Fabrication and Test of a Superconducting Dipole Magnet Based on Tilted

Solenoids," *IEEE Trans. Appl. Supercond. 17.2 (2007), pp. 2266-2269.*

[3] F. Bosi et al, "Compact Superconducting High Gradient Quadrupole Magnets for the Interaction Regions of High Luminosity Colliders," *IEEE Transactions on Applied Superconductivity, VOL. 23, NO. 3, JUNE 2013.*

[4] "The FCC-ee study," http://cern.ch/fcc .

[5] M. Koratzinos et al., "The FCC-ee Interaction Region Magnet Design," in *IPAC 2016*, Busan, Korea, THOPOR023.

[6] J. van Nugteren, "Internship Report: CERN, Software development for the Science and Design behind Superconducting Magnet Systems," tech. rep., Twente University: Energy Materials and Systems and CERN: ATLAS magnet team, 2011.